\documentclass[fleqn,usenatbib]{mnras}

\usepackage{newtxtext,newtxmath}
\usepackage[T1]{fontenc}
\usepackage{ae,aecompl}
\usepackage{graphicx}	
\usepackage{amsmath}	
\usepackage{amssymb}	
\usepackage[dvipsnames]{xcolor}
\usepackage{verbatim}
\usepackage{caption}
\usepackage{subcaption}
\usepackage{float}
\usepackage{breqn}
\newcommand{\nsources}{9270}
\newcommand{\figpath}{figures/}

\title[Differential accretion, MBHBs and the GWB]{The effect of differential accretion on the Gravitational Wave Background and the present day MBH Binary population}

\author[Magdalena S. Siwek et al.]{
Magdalena S. Siwek,$^{1}$\thanks{E-mail: magdalena.siwek@cfa.harvard.edu}
Luke Zoltan Kelley,$^{2}$
Lars Hernquist$^{1}$
\\
$^{1}$Center for Astrophysics, Harvard University, Cambridge, MA 02138, USA \\
$^{2}$Center for Interdisciplinary Exploration and Research in Astrophysics (CIERA), Northwestern University, Evanston, IL 60208, USA}

\date{Accepted XXX. Received YYY; in original form ZZZ}

\pubyear{2020}

\begin{document}
\label{firstpage}
\pagerange{\pageref{firstpage}--\pageref{lastpage}}
\maketitle

\begin{abstract}
Massive black hole binaries (MBHBs) form as a consequence of galaxy mergers. However, it is still unclear whether they typically merge within a Hubble time, and how accretion may affect their evolution. These questions will be addressed by pulsar timing arrays (PTAs), which aim to detect the GW background (GWB) emitted by MBHBs during the last Myrs of inspiral.
Here we investigate the influence of differential accretion on MBHB merger rates, chirp masses and the resulting GWB spectrum. We evolve a MBHB sample from the Illustris hydrodynamic cosmological simulation using semi-analytic models and for the first time self-consistently evolve their masses with binary accretion models. In all models, MBHBs coalesce with median total masses up to $1.5 \times 10^8 M_{\odot}$, up to $3-4$ times larger than in models neglecting accretion.  In our model with the largest plausible impact, the median mass ratio of coalescing MBHBs increases by a factor $3.6$, the coalescence rate by $52.3\%$, and the GWB amplitude by a factor $4.0$, yielding a dimensionless GWB strain  $A_{yr^{-1}} = 1 \times 10^{-15}$. Our model that favours accretion onto the primary MBH reduces the median mass ratio of coalescing MBHBs by a factor of $2.9$, and yields a GWB amplitude $A_{yr^{-1}} = 3.1 \times 10^{-16}$. This is nearly indistinguishable from our model neglecting accretion, despite higher MBHB masses at coalescence. \textbf{We further predict binary separation and mass ratio distributions of stalled MBHBs in the low-redshift universe, and find that these depend sensitively on binary accretion models. This presents the potential for combined EM and GW observational constraints on merger rates and accretion models of MBHB populations.}
\end{abstract}


\section{Introduction}

Almost every galaxy we observe harbors a supermassive black hole (MBH) at its centre \citep{Magorrian1998} and many galaxies in the local universe are expected to have undergone mergers throughout their evolution \citep{Lacey1994, Cole2002, RodriguezGomez2015}. As a consequence, one expects that at least some galaxies should contain multiple MBHs. Whether these MBHs migrate to the center of the remnant galaxy, form binaries and eventually merge, has been an active field of research for several decades (see e.g. \citealt{Merritt2005} for a review). The final stages of inspiral at sub-parsec separations are particularly poorly understood, and observationally almost unconstrained.

The evolution of MBHBs was first analytically explored in pioneering work by \cite{Begelman1980}. According to \cite{Begelman1980}, the evolution of the two MBHs in a remnant galaxy after merger is thought to start out at kpc scale separation, when the binary begins sinking down to the galactic center at timescales dictated by dynamical friction. When the MBHB separation has reduced to parsec-scales, the MBHs are immersed in the stellar background of the galactic bulge and will likely interact with individual stars \citep{Merritt2013}. Stars with sufficiently low angular momentum (the so-called "loss-cone"; LC) may scatter off the MBHB and carry some portion of angular momentum away from the central bulge  \citep{Shapiro1976}. This process is thought to shrink the orbital separation, or "harden" the binary, and may also modify the stellar background. MBHBs can likely only reach sub-parsec separations if enough stars remain available for scattering in the LC. If there are not enough stars in the core, depletion of the LC may lead to stalling of binaries at parsec-scales, known as the "final parsec problem" (see e.g. \citealt{Milosavljevic2003} for more details). However, if stellar scattering remains efficient due to re-filling of the LC (e.g. \citealt{Yu2002, Khan2011, Gualandris2016}), or if MBHBs are driven together by a third MBH (e.g. \citealt{Bonetti2016}), the binary may reach sub-parsec separation. At this point, if enough gas has been funneled into the center of the remnant galaxy (as seen in some simulations, e.g. \citealt{Barnes1991, Hernquist1991, Barnes1996,Hopkins2013}), a circumbinary accretion disk may form around the MBHB.

The interaction of circumbinary disks and binaries at scales ranging from T Tauri binary stars to MBHBs has been long studied (e.g. \citealt{Artymowicz1994}), and numerical simulations of such systems indicate that circumbinary disk accretion will push the mass ratio towards unity \citep{Bate2002, Farris2014, Gerosa2015, Duffell2019, Munoz2020}.
The masses and mass ratios of MBHBs, which emit gravitational waves (GWs) during the last Myrs of inspiral and produce the stochastic background of GWs (GWB; \citealt{Rajagopal1995, Phinney2001, Wyithe2003}), strongly affect the amplitude of the GWB.
Since MBHBs can evolve from galaxy merger until binary coalescence for Gyr timescales \citep{Kelley2016}, binary accretion may change the total masses and mass ratios of MBHBs significantly, and may have a potentially dramatic effect on the GWB amplitude.
Aside from facilitating accretion onto MBHBs, circumbinary disks are further believed to exert torques that may harden the binary separation at sub-parsec-scales until GW emission takes over (e.g. \citealt{Haiman2009}). At this final, GW-driven stage of inspiral, lasting a few Myrs, MBHBs emit most to their orbital energy in GWs, and build up the GWB spectrum. The fraction of binaries reaching this stage in a Hubble time is unknown, and crucially influences GWB amplitude predictions.

During MBHB evolution, accretion has been included in some studies that calculate the GWB, and accretion rates are either inferred from the remnant galaxy scaling relation (e.g. \citealt{Sesana2009, Rosado2015, Sesana2016}) or from semi-analytical models of the gas flow in the galactic bulge of each MBH (e.g. \citealt{Bonetti2018b}, based on the semi-analytic model by \citealt{Barausse2012}). In particular, \cite{Sesana2009} note that accretion onto both MBHs during inspiral may increase the GWB amplitude by a factor of $2$-$3$ by increasing the chirp masses of coalescing MBHBs. However, the effects of differential accretion on the MBHB population have not been studied in depth. Furthermore, no hydrodynamic circumbinary disk accretion models have been explored in the context of GWB predictions.
Our results are therefore unique, as \textbf{we apply self-consistently derived MBHB accretion rates from the Illustris simulations, and examine the impact of a comprehensive range of differential accretion models, including hydrodynamic circumbinary disk accretion models, on the MBHB population and the GWB}. 

As outlined here, the evolution of MBHBs at sub-parsec-scales is still poorly understood, and the observational side is no more complete.  Detections of MBHBs are currently only possible if both MBHs are luminous due to high accretion rates, and separated by large distances. Dual (gravitationally not bound) and binary AGN are two MBHs powered by accretion, which are both contained in their merged (or still merging) parent galaxies. Dual/binary AGN candidates are often identified by double-peaked emission lines (e.g. \citealt{Comerford2013}) or periodic variability in the optical/X-ray (e.g. \citealt{Graham2015}). Although many dual AGN at kpc separation have been found in the last two decades, only one binary AGN at parsec scale separation has been resolved \citep{Rodriguez2006}, and no confirmed sub-parsec binaries are known despite a growing pool of candidates (see \citealt{DeRosa2020} for a detailed review). However, we note the promising sub-parsec candidate OJ287, which has been observed for over a century and may contain an eccentric, very high-mass MBHB with a 12 year period (\citealt{Valtonen2008,Dey2018, Laine2020}).
Although there is still a dearth of definitive MBHB observations in the electromagnetic (EM) spectrum, a wealth of MBHB data will be available upon detection of the GWB in nHz frequencies.

Pulsar Timing Arrays (PTAs; \citealt{Foster1990}) are arrays of precisely timed millisecond pulsars. GWs affect the times of arrival (ToAs) of individual pulses, and PTA experiments calculate the difference between the expected and actual arrival time of successive pulses. For an array of very precisely timed millisecond pulsars, GWB signatures or upper limits of GWB amplitudes can be extracted from these timing residuals (see e.g. \citealt{Arzoumanian2018}).
Currently, 3 arrays of timed millisecond pulsars are in operation:
the European PTA (EPTA; \citealt{Desvignes2016}),
the Parkes PTA (PPTA; \citealt{Manchester2013}) and
North American Nanohertz Observatory for Gravitational Waves (NANOGrav; \citealt{NANOGrav2015})
which together form the International Pulsar Timing Array (IPTA; \citealt{Verbiest2016}). \cite{Taylor2016} predict a high ($\sim 80 \%$) probability that such large PTAs will detect GWs within this decade. Strong individual sources of GWs may either be detected well after the initial GWB detection \citep{Rosado2015}, or within a comparable time \citep{Kelley2017}.

Once detected, measurements of the GWB will confirm whether a significant fraction of MBHBs truly merge within a Hubble time. The GWB amplitude is sensitive to MBHB chirp masses and the cosmological merger rate of galaxies, while the spectral shape at low-frequencies will contain information about the sub-parsec-scale environmental effects that harden the binary separation before GW emission becomes effective.
To interpret the GWB signal correctly, we need to understand which key parameters during the MBHB evolution drive the amplitude and shape of the GWB most significantly. Here we investigate the impact of binary accretion models on the GWB and the MBHB population. While binary accretion can increase the chirp masses of coalescing MBHBs and boost the GWB amplitude, the cosmological galaxy merger rate and MBHB hardening efficiency also drives the GWB signal. For any given GWB measurement, it is therefore difficult to assess whether the signal originates from a smaller number of MBHBs with very high chirp masses, or many MBHBs with slightly lower chirp masses, introducing a degeneracy that will be difficult to break with GW measurements alone.
 Binary AGN at kpc-scales are already resolved in the EM spectrum, and in the course of this work we further show how binary accretion models may measurably affect the relative abundances of kpc-scale and parsec-scale MBHB populations. EM surveys of dual AGN/MBHBs and measurements of the GWB amplitude can therefore complement each other, break the degeneracy in the GWB strain measurement, and constrain accretion models as well as the galaxy merger rate which drives the MBHB formation rate.
Future multimessenger studies (such as outlined in \citealt{DeRosa2020, Kelley2020_whitepaper}) of MBHB populations are therefore likely to be essential to improving our understanding of MBHB formation, evolution and coalescence.

This paper is structured as follows. In section \ref{section:Illustris} we describe the sample of MBHBs we take from the Illustris simulation. Sections \ref{section:mbhbevol} and \ref{section:binaryacc} respectively explain the orbital hardening mechanisms and accretion models that evolve the binaries until coalescence.
In section \ref{sec:gwb} we show how we calculate the GWB spectrum emitted by a sample of MBHBs in a finite volume (as in the Illustris simulations). Results, Discussion and Conclusions are presented in sections \ref{sec:caveats}, \ref{section:results} and \ref{sec:conclusions}.

\section{Methods}
Here we describe the strategy we follow to select suitable MBHBs from Illustris (section \ref{section:Illustris}).
The construction of our MBHB sample from Illustris and their subsequent semi-analytic evolution (sections \ref{section:Illustris} and \ref{section:mbhbevol}) are based on the analysis first presented in \cite{Kelley2016, Kelley2017}, where more detailed information can be found.

\subsection{Illustris MBHBs}
\label{section:Illustris}
The Illustris simulations \citep{Vogelsberger2014a, Genel2014, Vogelsberger2014b} are a series of large-scale hydrodynamic simulations of galaxy formation, initialized at $z = 127$ in a cosmological box measuring 106.5 Mpc on each side.
As galaxies in Illustris merge, their respective MBH particles are placed into the remnant galaxy and are assumed merged at $\sim$ kpc separation \citep{Sijacki2015}.
In our model, the "merger" of two Illustris MBHs at near kpc-scales is treated as the \textit{formation} of a dual MBH system, which we evolve in semi-analytic post-processing models until coalescence or $z=0$.  Throughout the post-processing, we let the binaries accrete at rates consistent with their merged Illustris counterparts, such that the \textit{total mass of our post-processed binary is equal to the merged Illustris MBH at every subsequent timestep in the simulation.} We therefore arrive at the same BH masses as Illustris, which is in very good agreement with observations in the local universe \citep{Sijacki2015}.
From $z = 127$ to $z = 0$, Illustris records 23708 black hole merger events.
Following \cite{Kelley2016}, we apply mass cuts for MBHs ($M_{\bullet} \geq 10^6 M_{\odot}$) and require that their host galaxies are sufficiently well resolved. Out of the 23708 black hole merger events recorded in Illustris, we arrive at a fiducial sample of \nsources \ MBHBs that we evolve to coalescence.

\subsection{MBHB semi-analytic evolution}
\label{section:mbhbevol}
We follow the binary evolution prescription in \cite{Kelley2016} and \cite{Kelley2017}, which we summarize as follows. At kpc separation, all binaries form with a seed eccentricity $e_0 = 0.6$, and begin sinking towards the remnant core via dynamical friction (DF; \citealt{Begelman1980}). The hardening rate, or change in binary separation "a", takes the form,
\begin{equation}
\label{eqn:DF_hardening}
\frac{da}{dt} = - \frac{4 \pi G^2(M_2+M_{\star})\rho a}{v^3} \ln\Lambda ,
\end{equation}
where $M_2$ is the mass of the secondary MBH-galaxy system sinking towards the center of the merged remnant galaxy in a background of density $\rho$ with bodies of mass $M_{\star}$, and $v$ is the relative velocity of the secondary MBH and the background. $\ln\Lambda$ is the Coulomb Logarithm, quantifying the impact parameters of the scattering interaction between $M_2$ and the background.
DF is the dominant hardening mechanism down to $\sim 100$ parsec binary separation. From 100 to 1 parsec, binaries transfer most of their orbital energy through stellar Loss Cone (LC) scattering, which also increases the eccentricity of the orbit. Our LC treatment is based on the results from \cite{Sesana2006}, and we assume a full LC throughout our simulations.
Below parsec separations, the binaries are typically gravitationally bound and embedded in the gas-rich core of the remnant galaxy \citep{Barnes1991, Barnes1996, Mihos1996, Barnes2002}. At this stage, an accretion disk may form and interact with the binary via dynamical torques. We evolve our sample of MBHBs with the analytically derived disk torques in \cite{Haiman2009}. The circumbinary disk torques in our prescription reduce the orbital separation for all our binaries (see, however, more recent hydrodynamic simulations that observe binary expansion through disk torques, e.g.: \citealt{Miranda2016, Munoz2019b, Moody2019, Duffell2019, Munoz2020}), and thus remain the dominant hardening mechanism until the GW driven regime, which typically begins at $\sim 0.01$ parsec separation in our simulations. From this point onward until coalescence, the hardening rate of our binaries is dominated by the emission of GWs, which also circularizes the orbit and takes the form,
\begin{equation}
\label{eqn:Peters_dadt}
\frac{da}{dt} = -\frac{64G^3}{5c^5} \frac{M_1 M_2 (M_1 + M_2)}{a^3} \frac{(1 + \frac{73}{24}e^2+\frac{37}{96}e^4)}{(1-e^2)^{7/2}},
\end{equation}
following \cite{Peters1964}.

\subsection{Binary accretion models}
\label{section:binaryacc}

\begin{figure}
	\centering
	\includegraphics[width=\columnwidth]{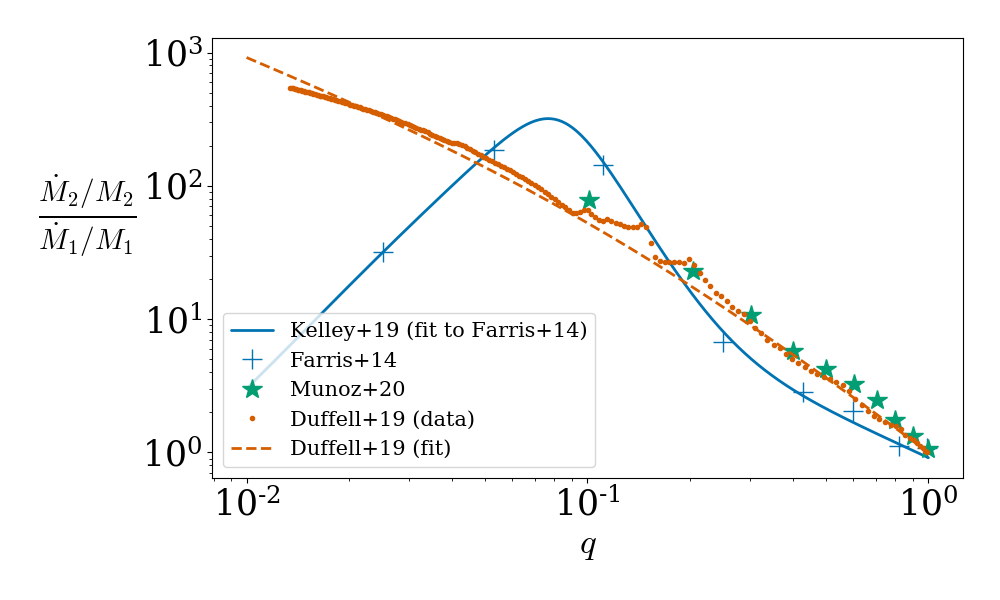}
	\caption{Ratio of relative accretion rates onto the secondary vs. the primary, as a function of mass ratio for several hydrodynamic simulations.
		Notably, the \protect\cite{Duffell2019} and \protect\cite{Munoz2020} simulations do not observe a turnover of the accretion rates at any mass ratio in the shown regime. When $\frac{\dot{M}_2/M_2}{\dot{M}_1/M_1} > 1$, the mass ratio of the binary increases.}
	\label{fig:relative_disk_accretion_comparison}
\end{figure}

Binary accretion is a highly non-linear problem and generally cannot be solved analytically.
Instead, hydrodynamic studies of circumbinary disks serve to study accretion rates and torques (see e.g. \citealt{Farris2014, Gerosa2015, Miranda2016, Moody2019, Duffell2019, Munoz2020}). In figure \ref{fig:relative_disk_accretion_comparison} we show a comparison of the relative accretion rates onto the secondary and primary MBH as found by some of the aforementioned studies.
As long as the accretion ratio $\frac{\dot{M}_2/M_2}{\dot{M}_1/M_1} > 1$, the mass ratio of the binary increases. In the results from \cite{Farris2014}, a turnover is seen at $q \sim 0.1$. The primary MBH will begin to accrete preferentially once the relative accretion ratio $\frac{\dot{M}_2/M_2}{\dot{M}_1/M_1} < 1$, which will decrease the mass ratios of binaries in this model.
However, the same turnover is not seen in other models. As we shall see, the existence of the turnover can affect the mass ratio distribution of the surviving MBHB population at $z = 0$ significantly.
Here we study the effect of binary accretion on MBHB mass ratios and the GWB. We further compare the effect of the hydrodynamic motivated binary accretion prescription from the densely sampled simulations by \cite{Duffell2019} to contrasting models.
Our accretion models are defined as follows.

\begin{itemize}
	\item \texttt{Mod\_NoAcc}: No accretion onto MBHBs post Illustris merger: $\dot{M} = 0$.
	\item \texttt{Mod\_Prim}: Accretion onto the primary only: $\dot{M}_1 = \dot{M}$, $\dot{M}_2 = 0$.
	\item \texttt{Mod\_Sec}: Accretion onto the secondary only: $\dot{M}_2 = \dot{M}$, $\dot{M}_1 = 0$.
	\item \texttt{Mod\_Prop}: Accretion onto each MBH proportional to its mass. We achieve this by splitting the total accretion rate $\dot{M}$ into individual MBH accretion rates as follows:
	\begin{equation}
		\dot{M}_1 = \frac{1}{1+q}\dot{M}, \ \ \ \dot{M}_2 = \frac{q}{1+q}\dot{M}, \ \ {\rm{where}} \ \ q = \frac{M_2}{M_1} \leq 1.
	\end{equation}
	This ensures a constant mass ratio throughout the binary evolution, but the same total mass growth of the binary in our models as the Illustris remnant MBH. This model allows us to distinguish the effects of MBH mass growth and mass ratio evolution on the GWB.
	\item \texttt{Mod\_Duffell}: Here we test the binary accretion prescription based on numerical simulations by \cite{Duffell2019}, which favours growth of the secondary MBH. In particular, we use their equation 8, relating the individual accretion rates to the mass ratio of the two MBHs as follows:

	\begin{equation}
	\label{eqn:Duffel_fit_eqn8}
		\frac{\dot{M}_2}{\dot{M}_1} = \frac{1}{0.1 + 0.9q}.
	\end{equation}
	\item \texttt{Mod\_Farris}: This model tests the binary accretion prescription based on numerical simulations by \cite{Farris2014}, which increases mass ratios for binaries with $q > 10^{-2}$. We primarily use this model to examine the impact of the relative accretion rate turnover (blue line in figure \ref{fig:relative_disk_accretion_comparison}) on persistent MBHB populations at $z = 0$.
	\item \texttt{Mod\_NoAccDuffell}: In this model we do not evolve MBHB masses for the majority of the evolution, from kpc-scales to parsec-scales. Once binary separations reach the final parsec, we invoke the accretion prescription by \cite{Duffell2019}. We use this model to test the significance of differential accretion during the final parsec only.
\end{itemize}

We note that the Illustris accretion rate is Eddington limited by the total mass of the Illustris remnant MBH. In our accretion models, the secondary MBH may accrete preferentially, and can thus exceed its own Eddington limit. During such phases, we reduce the secondary accretion rate to the Eddington limit, and add the excess to the primary MBH accretion rate instead. The Illustris remnant MBH accretion rates rarely approach the Eddington limit, and if so, only for short periods. Thus our treatment of this scenario has little effect on the final results.

\subsection{Calculating the GWB}
\label{sec:gwb}
The GWB spectrum was first calculated by \cite{Rajagopal1995} and has a simple dependence on the GW frequency \citep{Phinney2001}:
\begin{equation}
\label{eqn:gwbpropto}
h_c(f) \propto f^{-2/3},
\end{equation}
with an amplitude that depends on MBHB chirp masses and their total merger rate.
Accounting for eccentricity evolution, the GW energy spectrum takes the following form as derived by \citealt{Enoki2007}, equation 3.11:
\begin{equation}
\label{eqn:enoki3.11}
h_c^2(f) = \frac{4\pi c^3}{3} \int_{}^{}dz \int d\mathcal{M} \frac{d^2N}{d\mathcal{M} dz}  (1+z)^{-1/3} \Big( \frac{G\mathcal{M}}{c^3} \Big)^{5/3} (\pi f)^{-4/3} \Phi,
\end{equation}
where $\frac{d^2N}{d\mathcal{M} dz}$ is the comoving number density of GW sources per unit redshift interval $z \rightarrow z+dz$ and chirp mass interval $\mathcal{M} \rightarrow \mathcal{M} + d\mathcal{M}$. The chirp mass of a binary with combined mass $M = M_1 + M_2$ and and mass ratio $q$ is defined as follows:
\begin{align}
\label{chirpmass}
\mathcal{M} = \Bigg[ \frac{q}{(1+q)^2} \Bigg]^{3/5} \times M.
\end{align}
The final term, $\Phi$, accounts for eccentric orbits by distributing the total power of the GWs emitted by the binary among the harmonics of the orbital frequency:
\begin{equation}
\centering
\label{eqn:enoki3.123}
\Phi \equiv \sum_{n=1}^{\infty} \Big( \frac{2}{n} \Big)^{2/3} \frac{g(n,e)}{F(e)},
\end{equation}
where $g(n,e)$ is the GW frequency distribution function (see equation 2.4 in \citealt{Enoki2007} for details), and \mbox{$F(e) = \sum_{n=1}^{\infty} g(n,e) $}.

To construct the GWB from our sample of discrete Illustris MBHBs, we employ the Monte Carlo (MC) GWB calculation (e.g. \citealt{Sesana2008}). For each MBHB in our sample, this method counts the total number of individual GWB sources representative of that source in the observer's past light cone. In this work, we use the approach by \cite{Kelley2017}, which constructs probability distributions around each binary at each timestep of the Illustris simulation, and draws a likely number of similar sources from the observer's past light cone at the same redshift interval as the representative source.
In the MC method, the smooth MBHB distribution factor $\int_{}^{} dz d\mathcal{M} \frac{d^2N}{d\mathcal{M} dz}$ (equation \ref{eqn:enoki3.11}) is replaced by a factor that "weighs" each binary (denoted by index "i") at each timestep (index "j") by a factor $\Lambda_{ij} \equiv \frac{\Delta V_{ij}}{V_c}$.  $\Delta V_{ij}$ is the volume element of the past light cone that the binary lives in, and $V_c$ is the Illustris simulation box volume. Finally, the number of binaries represented by each member in our distribution is drawn from a Poisson distribution $\mathcal{P}$ centered around $\Lambda_{ij}$. In practice, this replaces the continuous MBHB distribution factor in equation \ref{eqn:enoki3.11} by integer numbers of sources:
\begin{equation}
	\int_{}{}dz d\mathcal{M}\, \frac{d^2 N }{dz d\mathcal{M}} \rightarrow \sum_{ij}^{} \mathcal{P} (\Lambda_{ij}).
\end{equation}

\section{Results}
\label{section:results}

\subsection{Total Mass Evolution}
\label{sec:mass_evol}
\begin{figure}
	\centering
	\includegraphics[width=\columnwidth]{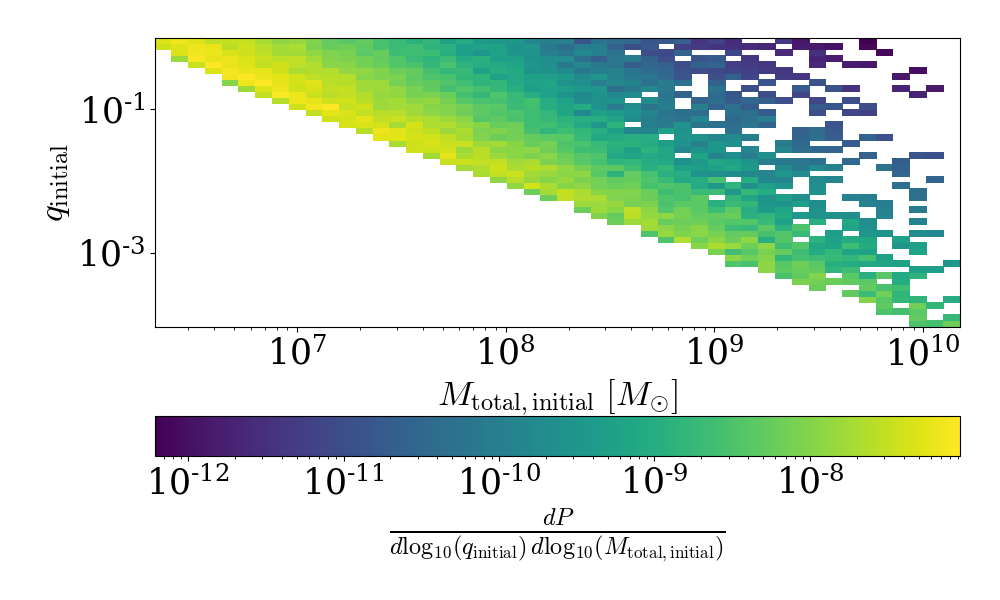}
	\caption{PDF of the initial mass ratios and initial total binary masses. In our sample, most binaries start out at low masses $M_{\bullet} \leq 10^7 M_{\odot}$ and high mass ratios. We attribute this to the abundance of small galaxies in the universe, making the formation of low mass, high mass ratio MBHBs more likely. Most binaries with high initial masses start out at low mass ratios, for the related reason that major mergers of massive galaxies are rare. We note that in our simulations, we consider only MBHs with masses above a lower limit of $10^6 M_{\odot}$, explaining why extremely low mass ratios are restricted. }
	\label{fig:q_vs_mtot}
\end{figure}
\begin{figure*}
	\includegraphics[width=\textwidth]{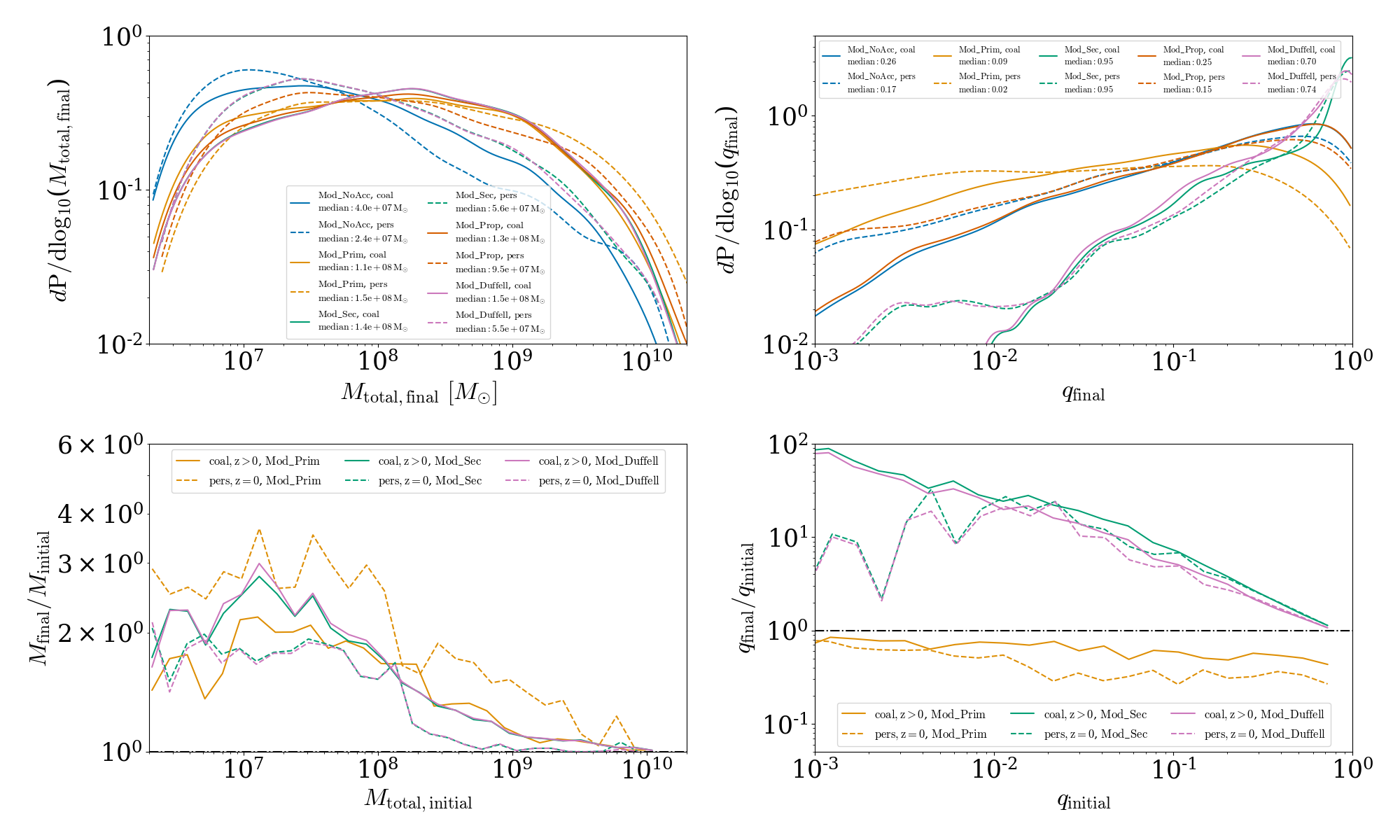}
	\caption{Top left: PDFs of final binary masses for coalesced (solid lines) and persisting (dashed lines) MBHBs. Coalesced binaries have median masses ranging from $4.0 \times 10^7 M_{\odot}$ (\texttt{Mod\_NoAcc}) to $1.5 \times 10^8 M_{\odot}$ (\texttt{Mod\_Duffell}). Persisting binaries have comparable median masses, ranging from $2.4 \times 10^7 M_{\odot}$ (\texttt{Mod\_NoAcc}) to $1.5 \times 10^8 M_{\odot}$ (\texttt{Mod\_Prim}). \\
		Top right: PDFs of final mass ratios for coalesced (solid lines) and persisting (dashed lines) MBHBs. Coalesced binaries have median mass ratios from $0.09$ (\texttt{Mod\_Prim}) to $0.95$ (\texttt{Mod\_Sec}), while persisting binaries have extremely low median mass ratios ($0.02$) in \texttt{Mod\_Prim} and again very high ($0.95$) median mass ratios in \texttt{Mod\_Sec}.\\
		Bottom left: Median mass increase per initial binary mass. Low-mass binaries increase their masses by median factors up to $\sim 4$, while high mass binaries see almost no total mass increase. There is no strong dependence on accretion models, however, persisting MBHBs in \texttt{Mod\_Prim} grow most in mass due to their long lifetimes (compare figure \ref{fig:lifetimes_median}). \\
		Bottom right: Median mass ratio increase per initial binary mass ratio. Coalescing binaries in models \texttt{Mod\_Sec} and \texttt{Mod\_Duffell} increase their mass ratios by up to two orders of magnitude if they start out at extremely low initial mass ratios. In \texttt{Mod\_Prim}, binaries decrease their mass ratios by a median half order of magnitude if their initial mass ratios are close to 1.}
	\label{fig:q_m_fi_hist}
\end{figure*}
\begin{figure}
	\centering
	\includegraphics[width=1\columnwidth]{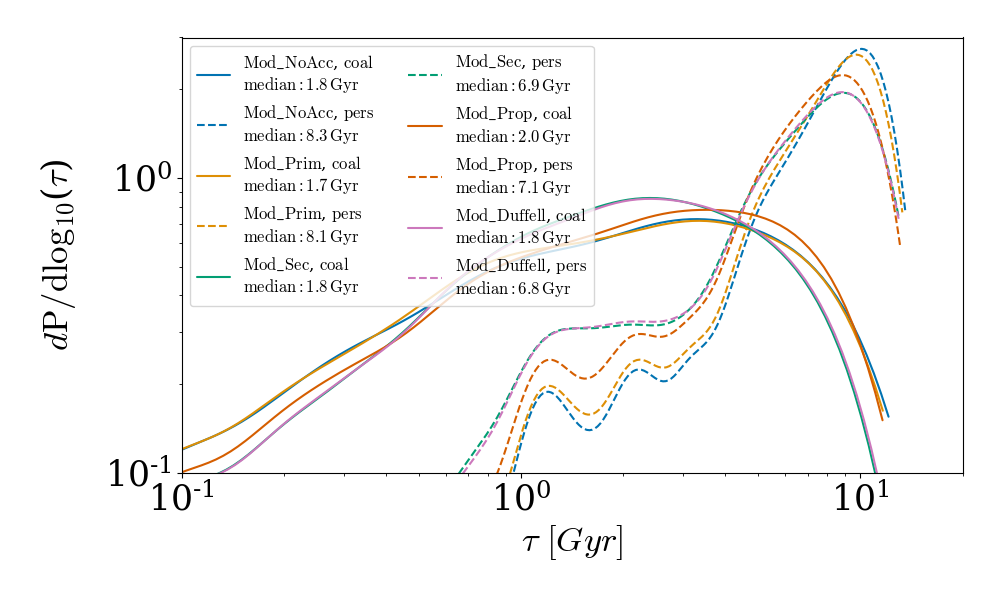}
	\caption{PDFs of lifetime for coalesced (solid lines) and persisting (dashed lines) binaries. The lifetimes here are defined as the time between Illustris merger and either coalescence in our models or until $z = 0$ is reached. We find that median lifetimes for coalesced binaries range from  $1.7$ Gyrs in \texttt{Mod\_Prim} to $2.0\,\rm Gyrs$ in \texttt{Mod\_Prop}. Persisting binaries are evolved for longer in our models, with medians ranging from $6.8\,\rm Gyrs$ in  \texttt{Mod\_Duffell} to $8.3\,\rm Gyrs$ in  \texttt{Mod\_NoAcc}.}
	\label{fig:lifetimes_median}
\end{figure}
Here we show the shift of the MBHB mass distribution in our simulations due to accretion during the Gyr timescales between galaxy merger and MBHB coalescence. The initial distribution of total binary masses is recorded at the times that Illustris assumes each binary "merged" (at $\sim$ kilo-parsec separations), and shown in figure \ref{fig:q_vs_mtot}, alongside the initial mass ratios at formation. Initial MBHB masses range from $2.0 \times 10^6\, M_{\odot}$ to $1.0 \times 10^{10} \, M_{\odot}$.
Final mass distributions of MBHBs that coalesce before $z = 0$ (the "coalescing fraction", solid lines), and final mass distributions of MBHBs that do not coalesce by $z = 0$ (the "persistent fraction", dashed lines) are shown in the top left panel of figure \ref{fig:q_m_fi_hist}.
The lifetime of a binary is defined as the time between Illustris MBHB "merger" and coalescence in our models. For persistent binaries, the lifetimes are defined as the time between Illustris MBHB "merger" and $z = 0$. Lifetimes are dependent on the accretion model, and in our coalescing binary fraction are shortest in \texttt{Mod\_Prim} and longest in \texttt{Mod\_Prop}  ($1.7$ and $2.0$ Gyrs, left panel in figure \ref{fig:lifetimes_median}), while persisting binaries are evolved to $z =0$ for medians of $6.8\,\rm Gyrs$ in  \texttt{Mod\_Duffell} to $8.3\,\rm Gyrs$ in \texttt{Mod\_NoAcc} (right panel in figure \ref{fig:lifetimes_median}). The median mass in the coalescing fraction is $4.0 \times 10^7 M_{\odot}$ in \texttt{Mod\_NoAcc} (top left panel in figure \ref{fig:q_m_fi_hist}). The shift of the median mass due to accretion in the coalescing fraction is smallest in \texttt{Mod\_Prim}, and largest in \texttt{Mod\_Duffell}. In the persisting fraction however, the opposite trend is seen: MBHBs in \texttt{Mod\_Duffell} have the smallest median mass shift, and MBHBs in \texttt{Mod\_Prim} have the largest median mass shift. This may indicate that more massive MBHBs are more likely to coalesce in \texttt{Mod\_Duffell} and stall in \texttt{Mod\_Prim}.
\textbf{We find that the medians of coalescing MBHB distributions shift by a factor of up to $\sim 4$ due to accretion during the binary evolution phase}.

\subsection{Mass Ratio Evolution}
\label{sec:q_evol}
We begin our simulations with a binary sample where small initial total masses correspond to the largest mass ratios (figure \ref{fig:q_vs_mtot}). Since all MBHs in our MBHB selection have masses $M_{\bullet} \geq 10^6 M_{\odot}$, very low mass ratios are restricted and low total mass binaries have initially high mass ratios. We further note that most high mass binaries start out with low mass ratios ($q \lesssim 10^{-1}$), as a result of the scarcity of major mergers for galaxies in this mass range.

The top right panel in figure \ref{fig:q_m_fi_hist} compares the final mass ratio distributions of the coalesced (solid lines) and persisting (dashed lines) populations. We find that in both MBHB populations, the median mass ratios are lowest in \texttt{Mod\_Prim} and highest in \texttt{Mod\_Sec}. \textbf{Comparing the mass ratios of coalescing and persisting populations in each model reveals that in most cases, the persisting populations have lower median mass ratios than coalesced MBHBs.}

The changes in mass ratios of coalescing and persisting MBHBs are shown in the bottom right panel of figure \ref{fig:q_m_fi_hist}. We only show \texttt{Mod\_Prim}, \texttt{Mod\_Duffell} and \texttt{Mod\_Sec}, as these are the only accretion models which impact the mass ratios.  We find that the lowest mass ratios in \texttt{Mod\_Duffell} and \texttt{Mod\_Sec} can increase from $q \sim 10^{-3}$ by up to a factor of $10^2$, whereas persisting binaries in the same models only increase from $10^{-3}$ by up to a factor of $10$.  \texttt{Mod\_Duffell} and \texttt{Mod\_Sec} have very similar effects on the mass ratio distribution. The secondary MBH in \texttt{Mod\_Duffell} preferentially accretes, and in \texttt{Mod\_Sec} accretes all of the available gas. This results in only a slightly higher mass ratio increase in \texttt{Mod\_Sec} versus \texttt{Mod\_Duffell}. As expected, MBHBs with mass ratios close to $\sim 1$ are barely affected in these models.
In \texttt{Mod\_Prim}, binaries with initially low mass ratios also do not evolve significantly from this initial value, however, the highest mass ratios with $q \lesssim 1$ can decrease by up to a factor of $\sim 3$.
Linking the initial mass ratios to the initial masses of the binaries, we observe that \texttt{Mod\_Duffell} and \texttt{Mod\_Sec} significantly boost the mass ratios of the \textbf{highest mass} MBHBs in our sample, whereas \texttt{Mod\_Prim} reduces the mass ratios of the \textbf{least massive} MBHBs. In \texttt{Mod\_NoAcc}, the median coalescing mass ratio is $0.26$, but depending on which MBH dominates the accretion, the median value across the coalescing MBHB population can decrease to $\sim0.09$ (\texttt{Mod\_Prim}) or increase to $\sim 0.95$ (\texttt{Mod\_Sec}).

\begin{figure}
	\centering
	\includegraphics[width=\columnwidth]{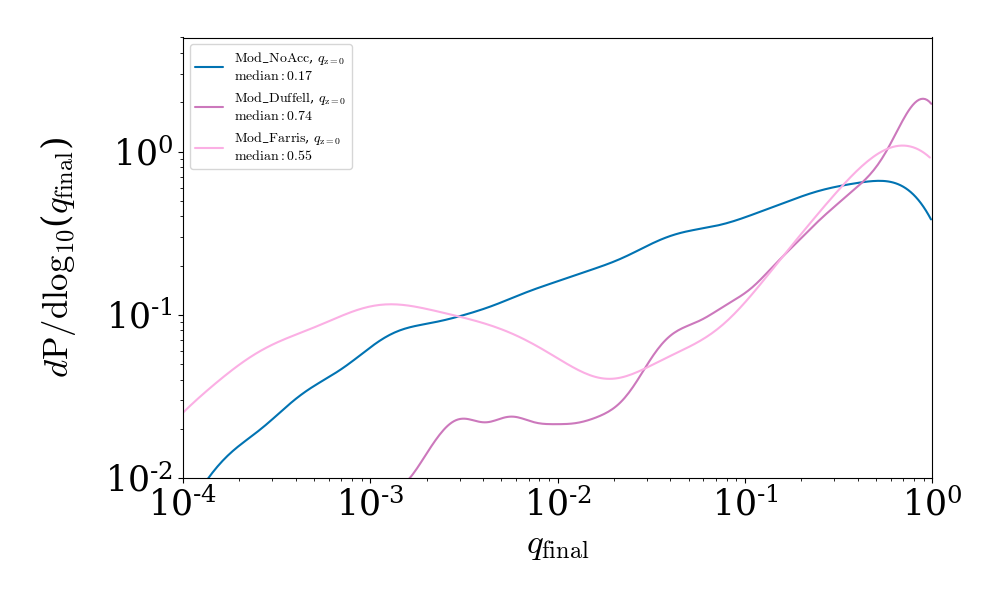}
	\caption{The mass ratio distributions of persisting binaries at redshift $z = 0$. The blue, purple and pink lines are the mass ratio distributions in \texttt{Mod\_NoAcc}, \texttt{Mod\_Duffell} and \texttt{Mod\_Farris} respectively at $z = 0$. The turnover in \texttt{Mod\_Farris} at $q \simeq 0.1$ evolves binaries with $q < 10^{-2}$ towards lower mass ratios, allowing a large population of low mass ratio binaries to survive until $z = 0$. The same is not the case in \texttt{Mod\_Duffell}, which does not see a turnover for the parameter range explored ($10^{-2} < q < 1$), driving all binaries towards higher mass ratios. Detections of extreme mass-ratio, very high-mass binaries at low redshift could suggest the presence of a turnover in the accretion partition function.}
	\label{fig:qf_nm}
\end{figure}

In figure \ref{fig:qf_nm} we compare the effects of the hydrodynamic derived binary accretion models from \cite{Farris2014} and \cite{Duffell2019} on the mass ratio distribution of persisting binaries at $z = 0$. These two models are representative of cases where accretion favours the secondary to varying degrees: \texttt{Mod\_Duffell} favours accretion onto the secondary for all mass ratios, whereas \texttt{Mod\_Farris} has a turnover at $q \sim 0.1$, and will begin to favour accretion onto the primary in binaries with very low mass ratios. In \texttt{Mod\_Farris} we see that the population of MBHBs in the present-day universe has been driven towards a bi-modal shape, with a depletion of binaries around $10^{-2} < q < 10^{-1}$. The relative accretion rates in the \cite{Farris2014} model drop below 1 at $q \lesssim 10^{-2}$, and thus start driving binaries in this parameter space to lower mass ratios, explaining the bimodal distribution in figure \ref{fig:qf_nm}.
In contrast, the \cite{Duffell2019} model has no turnover in the parameter space that was explored in their hydrodynamic models. We have therefore extrapolated their fitting function (see figure \ref{fig:relative_disk_accretion_comparison}) to lower mass ratios without a turnover. As a result, very few low mass ratio MBHBs persist in this model.
Relative accretion rates of binaries derived by hydrodynamic models likely should include a turnover at some low mass ratio, where the secondary mass becomes negligible compared to the primary mass. However, so far, only the sparsely sampled model by \cite{Farris2014} shows an indication of such a turnover in our mass ratio regime. The turnover mass ratio affects the mass ratio distribution of surviving MBHBs, presenting an opportunity to compare hydrodynamic binary accretion models with observations. 

\subsection{MBHB separations at z = 0}
\label{sec:mbhb_sep}
\begin{figure}
	\includegraphics[width=1\columnwidth]{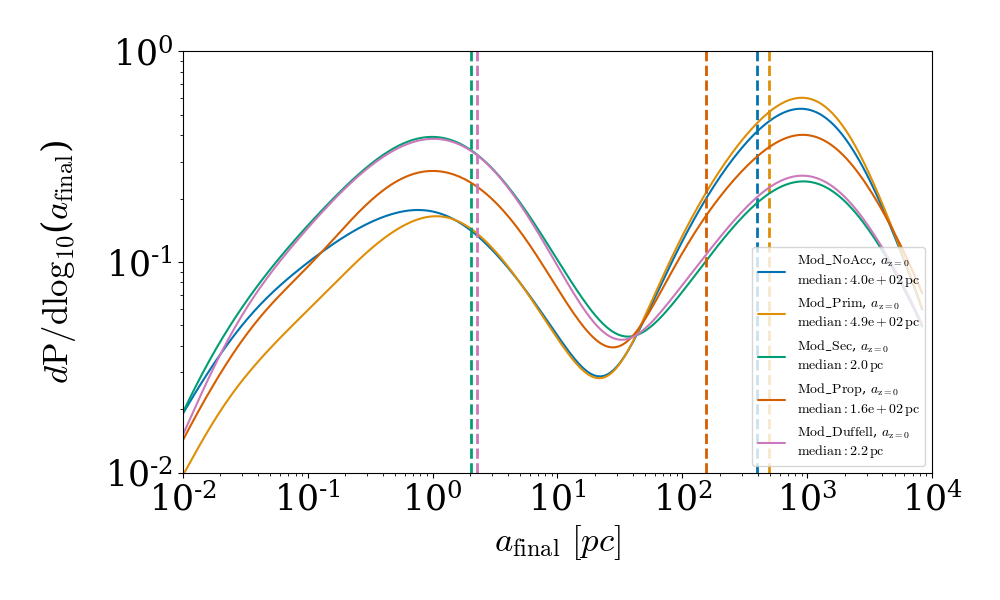}
	\caption{PDF of binary separation for all persistent MBHBs at $z = 0$. We find that there is a bi-modal distribution in the separation of persistent binaries, with peaks at $\sim 10^3$ and $\sim 1$ parsec. The relative amplitude of the peaks depends sensitively on the accretion model assumed during the evolution, thus providing a method to observationally constrain accretion during inspiral.}
	\label{fig:sep_hist}
\end{figure}

For our persisting population of MBHBs we find the distribution of binary separations at $z = 0$ in figure \ref{fig:sep_hist}. We show the PDF of binary separation for all accretion models, and find that each is bi-modal with peaks centred at $\sim 1 $ parsec and $\sim 1$ kpc. However, the peak at $\sim 1$ kpc is highest for \texttt{Mod\_Prim}, followed closely by \texttt{Mod\_NoAcc},  then \texttt{Mod\_Prop},  \texttt{Mod\_Duffell} and \texttt{Mod\_Sec}. The second peak at $\sim 1$ parsec is by half an order of magnitude highest for  \texttt{Mod\_Sec}, and lowest for \texttt{Mod\_Prim}, in exactly the reverse order as the first peak. In \texttt{Mod\_Duffell} and \texttt{Mod\_Sec}, more MBHBs stall at  $\sim 1 $ parsec than $\sim 1$ kpc, while the opposite is true in \texttt{Mod\_Prim}, \texttt{Mod\_NoAcc} and \texttt{Mod\_Prop}. This is likely caused by mass ratio dependent hardening rates in different regimes, and will be further explored in section \ref{sec:binary_hardening}.
 \textbf{The ratio of observed MBHBs at parsec vs. kpc separations appears to be indicative of the accretion model that evolved the population}. 

\subsection{Binary Hardening}
\label{sec:binary_hardening}
In figure \ref{fig:dadt_fractions_all} we show the median hardening timescales for DF, LC, VD and GW driven regimes, for all \textbf{coalescing} MBHBs in our model. The DF driven hardening timescale, $\tau_{DF}$, depends on the mass of the secondary MBH and its velocity (equation \ref{eqn:DF_hardening}), which depends inversely on the mass ratio.
The DF hardening timescale is lowest in \texttt{Mod\_Prim} and \texttt{Mod\_NoAcc} compared to all other models for the first $\sim 10^4$ parsec of binary evolution. We find that this is a selection effect for the coalesced population: \textbf{only binaries with initially higher mass ratios and secondary masses can coalesce by $z = 0$ if evolved in \texttt{Mod\_Prim} and \texttt{Mod\_NoAcc}, otherwise they will stall at kpc separation}. Those two models lead to more stalling of MBHB at kpc separation, and thus have lower coalescence rates.

The opposite effect is seen in models \texttt{Mod\_Sec} and \texttt{Mod\_Duffell}, where the DF driven hardening timescales are highest. 
In the coalescing population of MBHBs in these models, more low mass ratio, low secondary mass MBHs are evolved. The median DF hardening rate is initially longer due to the increased number of low mass ratio, low secondary mass MBHs that can coalescence in these models. As the secondary MBHs in these models accrete preferentially, their masses increase, driving $\tau_{DF}$ down. The preferential secondary accretion prevents these binaries from stalling in the kpc regime, facilitating coalescence before $z = 0$. 
Finally, while binaries evolved in \texttt{Mod\_Prop} do not change their mass ratios, secondary MBH masses are still grown through accretion. 
\textbf{The higher coalescing fractions in models \texttt{Mod\_Sec}, \texttt{Mod\_Duffell} and \texttt{Mod\_Prop} are caused by the secondary mass increase in these accretion models, which shortens the DF timescale and drives low secondary mass, low mass ratio binaries that would have otherwise stalled in the DF driven regime to coalescence.}

The LC and VD hardening rates do not differ significantly across our accretion models, however we note that the hierarchy in the timescales follows the dependence of the LC and VD regime on binary masses and mass ratios: in the same galactic environment, both circumbinary disk torques and LC scattering harden low mass, low mass ratio binaries most effectively. MBHBs in \texttt{Mod\_Prim} and \texttt{Mod\_NoAcc} therefore harden most efficiently, while \texttt{Mod\_Sec} and \texttt{Mod\_Duffell} have the longest hardening timescales in those regimes. Combining figures \ref{fig:dadt_fractions_all} and \ref{fig:sep_hist}, this explains the higher abundance of stalled MBHBs at parsec separations in \texttt{Mod\_Sec} and \texttt{Mod\_Duffell}. Similarly, the DF hardening dependence on mass ratio and secondary MBH mass explains the increased fraction of stalled MBHBs in \texttt{Mod\_Prim} and \texttt{Mod\_NoAcc} at kpc separations.

While the GW driven regime is negligibly short in comparison with the other regimes, we also point out that the GW timescales follow the hierarchy of median chirp masses in the coalescing fraction for each accretion model (see figure \ref{fig:mchirp_all}). 

\subsection{Gravitational Wave Background}
\begin{figure*}
	\centering
	\includegraphics[width=2\columnwidth]{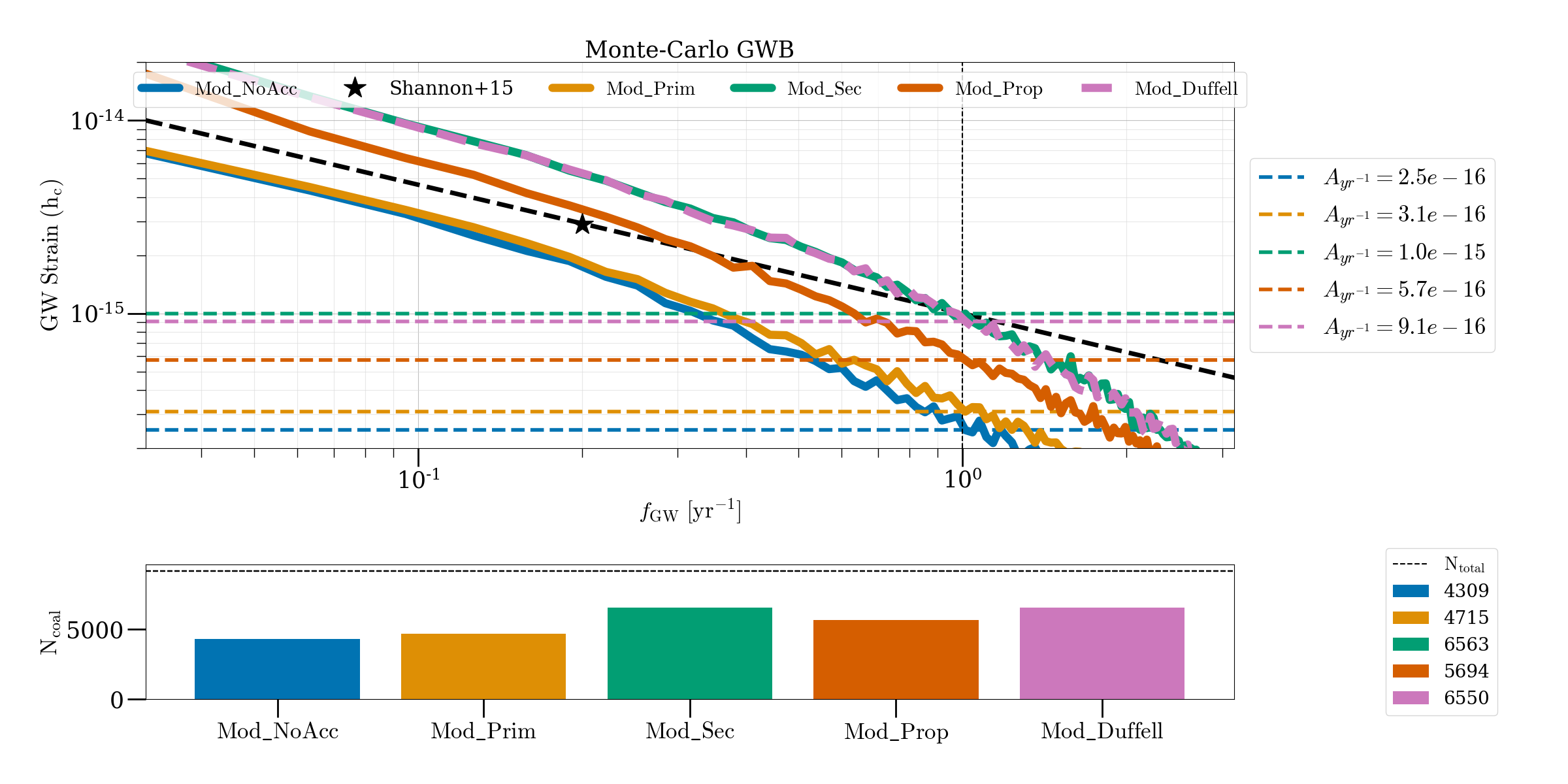}
	\caption{Top panel: MC GWB spectra. 
		\texttt{Mod\_Sec}, \texttt{Mod\_Prop} and \texttt{Mod\_Duffell} exceed the current PPTA sensitivity \citep{Shannon2015}, while \texttt{Mod\_NoAcc} and \texttt{Mod\_Prim} are below the current limit. While all accretion models increase the GWB strain, the factor by which the amplitude is boosted depends on which binary component grows. Accretion onto the secondary MBH has the largest effect (factor of $4$ increase) on the GWB amplitude. Even our most conservative model (\texttt{Mod\_NoAccDuffell}; shown in figure \ref{fig:gwb_noacc_duffell}), where the \protect\cite{Duffell2019} accretion model is applied only during the last parsec, shows a significant ($68\%$) increase in the GWB amplitude. \\
	Bottom panel: Number of mergers in our MBHB sample per model. We find that out of \nsources \ systems, between  46.5\% (\texttt{Mod\_NoAcc}) and 70.8\% (\texttt{Mod\_Sec}) of MBHBs coalesce in a Hubble time.}
	\label{fig:gwb}
\end{figure*}

We now present the GWB spectra and MBHB coalescence rates for each of our models in figure \ref{fig:gwb}. We compare \texttt{Mod\_Prim}, \texttt{Mod\_Sec}, \texttt{Mod\_Prop} and \texttt{Mod\_Duffell} to \texttt{Mod\_NoAcc}, where binaries are evolved without accretion.
We find that GWB amplitudes at the frequency $f = 1 \rm yr^{-1}$ are ranked depending on the accretion model as follows:
\begin{equation}
\label{eqn:gwb_hierarchy}
\begin{aligned}
	A_{yr^{-1}, \texttt{Mod\_NoAcc}} \lesssim A_{yr^{-1}, \texttt{Mod\_Prim}} < A_{yr^{-1}, \texttt{Mod\_Prop}} \\
	 < A_{yr^{-1}, \texttt{Mod\_Duffell}} \lesssim A_{yr^{-1}, \texttt{Mod\_Sec}},
\end{aligned}
\end{equation}
and have numerical values shown in figure \ref{fig:gwb}. 
We find that \texttt{Mod\_Prim} shows only a small increase in its GWB amplitude (24.0\%) and coalescence rate (9.4\%) compared to \texttt{Mod\_NoAcc}, despite the increased total masses of coalescing MBHBs in this model. We attribute this to the decreased median mass ratios, and thus lower chirp masses of coalescing binaries in this model.
\texttt{Mod\_Prop}, which allows accretion onto both MBHs at rates proportional to their masses, results in a more significant GWB amplitude increase by a factor of $2.3$, and a merger rate increase of 32.1\%. The GWB amplitude increase we see in \texttt{Mod\_Prop} is consistent with \cite{Sesana2009}, who apply a similar accretion model and find GWB amplitude increases by a factor $2$ - $3$. While mass ratios in \texttt{Mod\_Prop} are not evolved, the total binary masses increase and thus the median chirp mass of the coalescing binaries in the GWB relevant mass regime have increased, boosting the amplitude of the GWB.
The most significant increase in the GWB amplitudes and merger rates is seen in models \texttt{Mod\_Sec} and \texttt{Mod\_Duffell}. Both models favour the secondary MBH in their accretion prescription, which increases median chirp masses in the coalescing binary fractions. The coalescence rates in \texttt{Mod\_Sec} and \texttt{Mod\_Duffell} are increased by $52.3 \%$ and $52.0 \%$, and the GWB amplitudes by a factor $4.0$ and $3.6$ respectively, compared to \texttt{Mod\_NoAcc}.
In figure \ref{fig:mchirp_all} we show the chirp mass distributions of coalescing MBHBs in our sample. We find that the median chirp masses of coalescing MBHBs in \texttt{Mod\_NoAcc} are smallest, and those in \texttt{Mod\_Sec} are largest, following the same hierarchy as the GWB amplitudes in equation \ref{eqn:gwb_hierarchy}.

To verify the importance of binary accretion models in the final parsec only, we also tested a scenario where MBHBs do not accrete \textit{until the final parsec}, and then accrete following the \cite{Duffell2019} prescription (\texttt{Mod\_NoAccDuffell}). We present the findings in figure \ref{fig:gwb_noacc_duffell} and find that \textbf{allowing for binary accretion in the final parsec only still yields a GWB amplitude increase of $68.0\%$}, while coalescence rates increase by $7.8\%$.

\subsubsection{Analytical Comparisons}
\label{sec:gwb_analytical}
The way mass was added to the binaries in our accretion models has affected the amplitude boost of the resulting GWB. Here we give a brief analytical justification of the amplitude boosts in models (\texttt{Mod\_Prim}, \texttt{Mod\_Sec} and \texttt{Mod\_Prop}) as compared to the baseline amplitude resulting from \texttt{Mod\_NoAcc}. We do this by calculating GWB amplitude boosting factors that simply depend on the change in total mass and mass ratio in each model, and estimating the median boost from all coalescing binaries.

In \texttt{Mod\_Prim}, mass is added only to the primary, increasing the total mass but decreasing the mass ratio. By setting the binary mass $M \rightarrow M + \delta M$ and the mass ratio $q = \frac{M_2}{M_1} \rightarrow \frac{M_2}{M_1 + \delta M} $, it is easy to show that the amplitude of the GWB, $h_c^2$, (see equation \ref{eqn:enoki3.11}) is boosted by a factor dependent on $\delta M$ to give the accretion-modulated amplitude $h_{\rm c, mod}^2$:

\begin{equation}
\label{eqn:hcdash_Mod2}
	h_{\rm c, mod}^2 = h_c^2 \times \frac{1 + \delta M/M_1}{(1+\delta M/M)^{1/3}}
\end{equation}

In \texttt{Mod\_Sec}, we follow the same strategy, setting the binary mass $M \rightarrow M + \delta M$ and the mass ratio $q = \frac{M_2}{M_1} \rightarrow \frac{M_2 + \delta M}{M_1} $.
We find the amplitude of the GWB to be modulated as follows:

\begin{equation}
\label{eqn:hcdash_Mod3}
h_{\rm c, mod}^2 = h_c^2 \times \frac{1 + \delta M/M}{(1+\delta M/M)^{1/3}}.
\end{equation}

Finally, in \texttt{Mod\_Prop}, where the MBH accretion rates are proportional to the initial mass ratio, only the binary mass is changed $M \rightarrow M + \delta M$, and yields the amplitude modulation:

\begin{equation}
\label{eqn:hcdash_Mod4}
h_{\rm c, mod}^2 = h_c^2 \times (1 + \delta M/M)^{5/3}.
\end{equation}
From the top left and top right panels of figure \ref{fig:q_m_fi_hist}, setting the median mass increase $\delta M/M \sim 3$ and initial median mass ratio $q \sim 0.2$, we recover the hierarchy of the GWB amplitudes we find in figure \ref{fig:gwb}:
\begin{equation}
h_{\rm c, mod}^2 (\texttt{Mod\_Prim})  < h_{\rm c, mod}^2 (\texttt{Mod\_Prop}) \lesssim h_{\rm c, mod}^2 (\texttt{Mod\_Sec}).
\end{equation}

These simple analytic exercises provide intuition of the GWB amplitude hierarchy in our models.

\section{Discussion}
\label{sec:caveats}
Here we discuss our results and point out caveats that should be considered when interpreting our findings. We point out future research directions that can be explored based on our work.
\begin{enumerate}
	\item In agreement with our initial hypothesis we have found that binary accretion models with high secondary MBH accretion rates significantly boost the GWB amplitude. These models also increase the MBHB coalescence rates, which we attribute to higher DF hardening rates in binaries with initially low secondary masses, that would have otherwise stalled in the DF regime at kpc separations.
	The GWB amplitudes in these models (\texttt{Mod\_Sec}, \texttt{Mod\_Prop} and \texttt{Mod\_Duffell}) exceed the current PPTA sensitivity. One possible solution for this conflict is that these models should be ruled out, and that the primary MBH accretes the majority of the available gas. Other solutions could include uncertainties in the cosmological galaxy merger rate, efficiency of LC scattering, sign of disk torques during the VD phase or other aspects of our idealized simulations.
	\item The GWB amplitudes in our models without accretion (\texttt{Mod\_NoAcc}) or accretion on the primary MBH only (\texttt{Mod\_Prim}) are comparable to similar MC GWB amplitude predictions (e.g. \citealt{Kelley2017}). The similar amplitudes in \texttt{Mod\_NoAcc} and \texttt{Mod\_Prim} indicate that if the primary MBH accretes most of the available gas, GWB measurements will likely not be able to distinguish between these two accretion models. However, we predict that EM observations should reveal an excess of very low mass ratio persisting MBHBs if \texttt{Mod\_Prim} is representative of true MBHB accretion behaviour.
	\item Following \cite{Haiman2009} we evolved our MBHBs with disk torques that harden the binary separation. More recent numerical simulations find that disk torques can be positive for a wide range of binary parameters, and may increase the binary separation (see e.g. \citealt{Miranda2016, Munoz2019b, Moody2019, Duffell2019, Munoz2020}, although \cite{Tiede2020} find that torques become negative for smaller disk scale heights). However, circumbinary disk simulations still make simplifying assumptions, for instance 2D co-planar treatment of the disk and binary or fixed viscosities. It is therefore still unclear how circumbinary disk torques are going to affect the MBHB coalescence rate and GWB.
	If numerical simulations for a wider parameter space (varying both disk inclination angles and mass ratios, for instance) become available, our work could be adapted to include circumbinary disk torque prescriptions motivated by numerical simulations, which may expand the orbital separations of some MBHBs in our sample, affecting coalescence rates and GWB amplitudes.
	\item By applying results of circumbinary disk simulations in \texttt{Mod\_Duffell} and \texttt{Mod\_Farris} from $10^4$ parsec until coalescence, we have predicted potentially observable features in present-day stalled MBHB populations. \textbf{With a statistical sample of MBHBs in the present-day universe, the abundance of very low mass ratio MBHBs could place constraints on preferential accretion rates in binaries and the existence of a low mass ratio turnover.}
	However, these models are contingent on a disk forming around a closed binary, which is unlikely at separations exceeding $\sim 1$ parsec. Due to the uncertainty around the true duration of the VD phase, we tested these models for the entire MBHB lifetime to provide upper limits for binaries that may be stalled in this phase. We have also verified that if disk accretion first starts at binary separations below $\sim 1$ parsec, with no accretion taking place until then, the GWB amplitude still increases by $68.0\%$, indicating that \textbf{even if preferential secondary accretion only takes place in the final parsec of inspiral, the GWB amplitude is still affected significantly.} Future work could include more detailed treatment of the accretion behaviour of MBHBs at kpc to parsec separations, potentially constrained by observations of dual AGN at these scales. High resolution EM observations of nearby sub-parsec MBHBs, such as the promising candidate OJ287 \citep{Valtonen2008}, could further verify the validity of the hydrodynamic sub-parsec circumbinary disk accretion models we test here.
	\item Due to the idealized nature of our models, we view the evolution of our MBHB sample with contrasting binary accretion models as an experiment. The resulting GWB spectra and MBH population properties we predict are meant to bracket the true result, rather than being taken at face-value. For instance, measurements of low GWB amplitudes could indicate that primary accretion is favoured during a significant fraction of MBHB evolution, while a high GWB amplitude could indicate that the secondary accretes at higher rates. However, EM observations of stalled MBHB populations are likely needed to place additional constraints on the binary accretion model, and help break the degeneracy between modified chirp mass distributions and MBHB merger rates.
	\item We have predicted binary separation distributions of stalled MBHBs, and found that most MBHBs in the local universe are either at $\sim 1$ parsec or $\sim 1$ kpc separation. In particular, we found that \textbf{the relative density of dual AGN at kpc and MBHBs at parsec separations scales with the binary accretion model following the galaxy merger, and could be directly compared to existing dual AGN observations at kpc separations.}
	 However, we point out that it is still difficult to confirm MBHB candidates at kpc and especially parsec separations, posing the risk of biases in the relative abundances of stalled MBHBs at $\sim 1$ parsec versus $\sim 1$ kpc separations.
	\item The role of the Eddington limit and radiative feedback is unclear in circumbinary disk accretion. We limit the accretion rate onto the secondary MBH by its own Eddington limit. Should $\dot{M}_2$ exceed this limit, we feed the excess into the primary MBH to allow for total binary mass growth consistent with the merged Illustris MBH.
	While the possibility of super-Eddington accretion in a binary should not be excluded, this prescription mimics radiative feedback of the secondary MBH, which may launch a wind to expel inflowing gas during super-Eddington accretion episodes. The gas that is expelled through the wind may then fall back onto the primary MBH.
	In Illustris, the accretion rates are limited by the Eddington limit of the total mass of the MBHB, and rarely approach this limit. Therefore our results are not significantly affected by this treatment.
	However, an intriguing aspect of radiative feedback in this context is the transport of angular momentum through winds from the individual accretion disks around each MBH. If a significant amount of angular momentum can be expelled from the binary in this way, radiative feedback could act as a new channel for MBHB hardening at parsec-scales.
	\item MBHB eccentricities were evolved through stellar scattering interactions (increasing eccentricity) and GW emission (decreasing eccentricity) in our model. All MBHBs were initialized with a seed eccentricity $e_0 = 0.6$. While MBHBs likely form with varying initial eccentricities, the value we chose is representative of a moderate seed eccentricity.  \cite{Kelley2017} show in their figure 7 that the choice of seed eccentricity has only a small effect on the final GWB amplitude, unless MBHBs are initialized with extremely high eccentricities ($e_0 = 0.99$), which would flatten the GWB spectrum.
	\item We have not included the possible formation of MBH triples (e.g. \citealt{Hoffman2007}) through repeated galaxy mergers in our model. Kozai-Lidov oscillations \citep{Kozai1962, Lidov1962} induced by a third MBH could increase the coalescing fraction of MBHBs (e.g.  \citealt{Omer2002, Bonetti2018a, Ryu2018}), and potentially boost the predicted GWB amplitude. Future extensions of this work could consider the formation of MBH triples in the Illustris simulations and examine the likely effect on the GWB amplitude.
\end{enumerate}

\section{Summary and Conclusions}
\label{sec:conclusions}
We have evolved a sample of \nsources \ MBHBs from the Illustris cosmological simulation from kpc separations until coalescense, using semi-analytic models including a variety of binary accretion models. We have investigated the effects of the accretion models on MBHB mass ratios, total masses, lifetimes, merger rates and the resulting GWB amplitudes. We have further given observable signatures of our accretion models in the population of persisting MBHBs at $z = 0$, including their mass ratio and separation distributions.

We find that regardless of accretion model, the lifetimes of coalescing binaries are of similar order $1 - 2$ Gyr (figure \ref{fig:lifetimes_median}), and allow for enough time to increase the total binary masses by factors of $\sim 3-4$ (top left panel in figure \ref{fig:q_m_fi_hist}). These timescales are even more significant for MBHBs that do not coalesce in a Hubble time (the "persisting population"; figure \ref{fig:lifetimes_median}).
We therefore conclude that \textbf{due to the long timescales between galaxy merger and MBHB coalescence, accretion is a non-negligible factor and must be considered to construct a realistic population of coalescing and persisting MBHBs}.
The binary accretion models we test here have varying impact on the mass ratio distributions: \texttt{Mod\_NoAcc} includes no accretion at all, \texttt{Mod\_Prim} accretes onto the primary only, decreasing the mass ratio and \texttt{Mod\_Sec} accretes onto the secondary only, increasing the mass ratio. \texttt{Mod\_Prop} has no effect on the mass ratio but grows each MBH proportional to its initial mass and \texttt{Mod\_Duffell} is a hydrodynamic motivated model which favours accretion onto the secondary.
Compared to the median initial mass ratio value of $0.26$ in \texttt{Mod\_NoAcc} (top right panel in figure \ref{fig:q_m_fi_hist}), we find that the coalescing MBHB population shows reduced mass ratios in \texttt{Mod\_Prim} ($0.09$, solid yellow line in top right panel in figure \ref{fig:q_m_fi_hist}), and increased mass ratios up to $0.95$ in \texttt{Mod\_Sec}.

We further find that the present-day mass ratio distribution of persisting MBHB can be sensitively affected by low mass ratio accretion behaviour. The properties of the relative accretion prescription, particular the mass ratio at which the secondary becomes the main accretor in the system, could then leave an imprint in the distribution of persisting MBHBs at $z = 0$. This distribution may be bi-modal, with very few MBHBs around the mass ratio associated with the turnover $\frac{\dot{M}_2/M_2}{\dot{M}_1/M_1}$  (see figure \ref{fig:qf_nm}), and an excess of stalled MBHBs with very low mass ratios.
Similarly, the distribution of dual AGN/MBHB separations depends on the accretion model (figure \ref{fig:sep_hist}), where accretion models that tend to increase mass ratios significantly boost the present number of MBHBs stalled at $\sim$ parsec separations. Accretion models which lower the mass ratio or keep it constant result in an excess of kpc separation dual MBHs, and fewer parsec scale MBHBs. On the other hand, models which favour accretion onto the secondary MBH result in more stalled MBHBs at parsec separations. We conclude that \textbf{MBHB accretion models may be constrained by the relative abundance of low-redshift kpc-scale dual AGN and parsec-scale MBHBs}.

We have further predicted the GWB amplitude and coalescence rates resulting from our MBHB populations. We found that in \texttt{Mod\_Prim}, the GWB amplitude at $f = 1 \rm yr^{-1}$ was increased by $24.0\%$ with respect to that resulting from \texttt{Mod\_NoAcc}. GWB spectra from MBHB populations evolved with \texttt{Mod\_Prim} are thus observationally hard to distinguish from populations that do not accrete any gas prior to coalescence. We conclude that this is due to lowered chirp masses in the coalescing fraction of \texttt{Mod\_Prim}. All other models, including \texttt{Mod\_Prop}, \texttt{Mod\_Sec} and \texttt{Mod\_Duffell} significantly boost the GWB amplitude by up to a factor of $\sim 4$ (figure \ref{fig:gwb}), due to their higher chirp masses at coalescence and higher coalescence rates. \textbf{By contrasting the mass ratio increasing (\texttt{Mod\_Sec} and \texttt{Mod\_Duffell}) and mass ratio decreasing (\texttt{Mod\_Prim}) accretion models, we conclude that mass ratio evolution, which is governed by the relative accretion of the two MBHs in the remnant galaxy, plays an even more significant role than total mass evolution of the binary with respect to MBHB coalescence rates and GWB amplitude.}
We have also considered a case where MBHBs are evolved without accretion until parsec separations, at which point the hydrodynamic circumbinary disk accretion model by \cite{Duffell2019} is invoked. In this case, the GWB amplitude increases by $68.0\%$ (figure \ref{fig:gwb_noacc_duffell}), indicating that \textbf{circumbinary accretion models in the final parsec of inspiral can be significant, despite the possibly short duration of this phase.}

To develop accurate predictions for GWB spectra and stalled MBHB populations it is crucial to understand the dynamics and growth of MBHBs in their remnant galaxies. In this study we have found that despite the lack of sub-parsec MBHB observations and associated EM signatures of accretion, \textbf{population studies of kpc scale dual AGN and measurements of the GWB amplitude can constrain the accretion behaviour of MBHs in their remnant galaxies}. In particular, EM observations of kpc scale dual AGN and independent GWB observations may help break the degeneracy between MBHB merger rates and accretion-boosted chirp mass distributions, which both affect the GWB amplitude. In this study we presented GWB amplitudes for a wide variety of binary accretion models, and model-dependent properties of stalled MBHB/dual AGN populations for this purpose. Together, GWB signatures and EM observations may help decode the population properties of coalesced and stalled MBHBs in the high and low redshift universe.

\section*{Acknowledgments}
We are grateful to Paul Duffell, Yuri Levin and Chiara Mingarelli for helpful discussions, especially during the early stages of this project.
This research made use of \texttt{SciPy} \citep{2020SciPy-NMeth} and \texttt{NumPy} \citep{vanderWalt2011}.
\texttt{Seaborn} \citep{seaborn} and \texttt{MATPLOTLIB} \citep{Hunter2007} were used to generate figures.

\bibliographystyle{mnras}
\bibliography{mybib} 

\appendix
\section{Appendix}

\begin{figure*}
	\centering
	\includegraphics[width=1\textwidth]{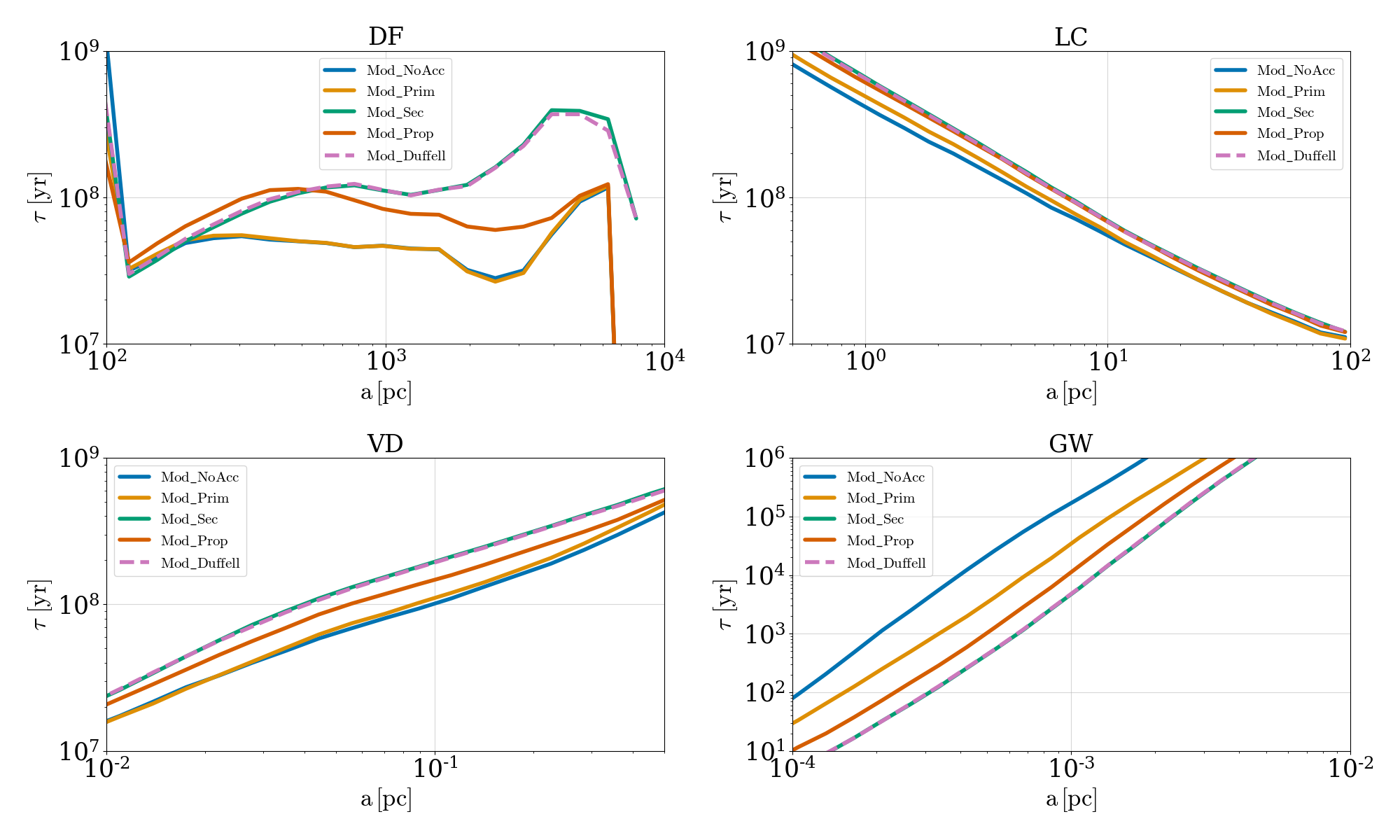}
	\caption{
		Top left: The median DF driven hardening timescale of coalescing binaries as a function of binary separation for our suite of accretion models. Since the DF hardening timescale is limited by the timescale at which the secondary MBH sinks towards the center of the galaxy, models with equal or similar secondary masses (\texttt{Mod\_NoAcc} and \texttt{Mod\_Prim}, \texttt{Mod\_Sec} and \texttt{Mod\_Duffell}) are similar.
		As the binaries approach $10^2$ parsec separations, DF is attenuated, explaining the sharp increase in hardening timescales. This marks the beginning of the LC driven hardening phase.\\
		Top right: The median stellar scattering / LC driven hardening timescales of coalescing binaries as a function of binary separation for our suite of accretion models. We find that binaries evolved without any accretion are hardened fastest in this regime. Accretion models that increase or decrease the mass ratio boost the LC timescale by roughly the same amount. \texttt{Mod\_Prop}, which grows the binary mass without altering the mass ratio shows the largest increase in the LC timescale. \\
		Bottom left: The median circumbinary disk driven hardening timescale of coalescing binaries as a function of binary separation for our suite of accretion models. Circumbinary disk torques scale inversely with binary mass ratios and total masses, so \texttt{Mod\_NoAcc} and \texttt{Mod\_Prim} are experiencing the largest torques, and thus shortest hardening timescales. \texttt{Mod\_Sec} and \texttt{Mod\_Duffell} have large mass ratios, and thus are least affected by ciucumbinary disk torques. \\
		Bottom right: Median GW driven hardening timescale of coalescing binaries as a function of binary separation for our suite of accretion models. We find that \texttt{Mod\_NoAcc} evolved binaries have the longest GW hardening timescales, followed by \texttt{Mod\_Prim}, \texttt{Mod\_Prop}, \texttt{Mod\_Sec}, and \texttt{Mod\_Duffell} with the shortest timescales. }
	\label{fig:dadt_fractions_all}
\end{figure*}

\begin{figure}
	\centering
	\includegraphics[width=1\columnwidth]{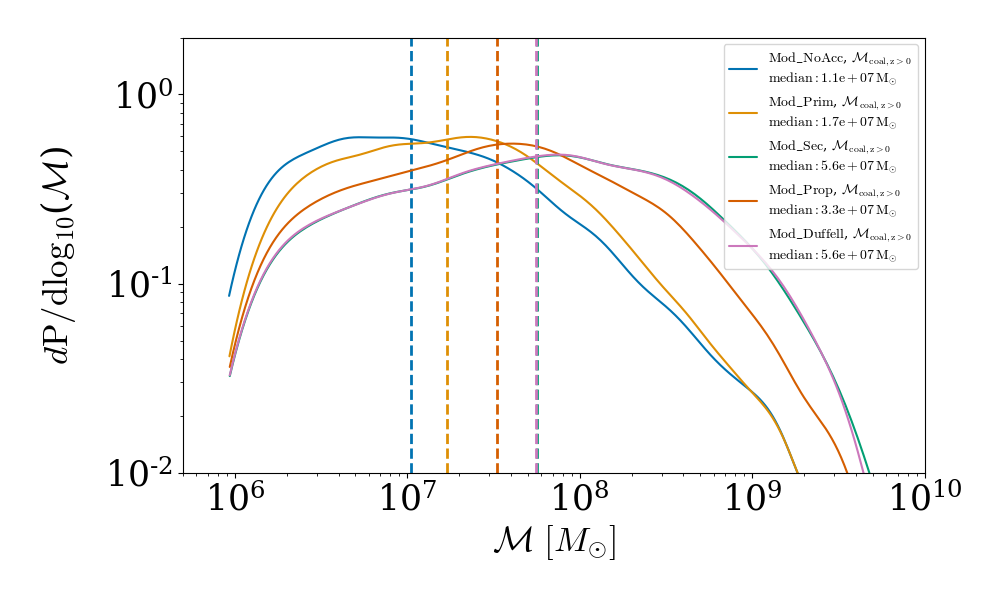}
	\caption{Chirp masses in the coalescing fraction for our accretion models. MBHBs in \texttt{Mod\_NoAcc} coalesce with the lowest chirp masses (median $1.1 \times 10^7 M_{\odot}$), followed by \texttt{Mod\_Prim} (median $1.7 \times 10^7 M_{\odot}$), \texttt{Mod\_Prop} (median $3.3 \times 10^7 M_{\odot}$), \texttt{Mod\_Sec} (median $5.6 \times 10^7 M_{\odot}$), and \texttt{Mod\_Duffell} (median $5.6 \times 10^7 M_{\odot}$).}
	\label{fig:mchirp_all}
\end{figure}

\begin{figure}
	\centering
	\includegraphics[width=1\columnwidth]{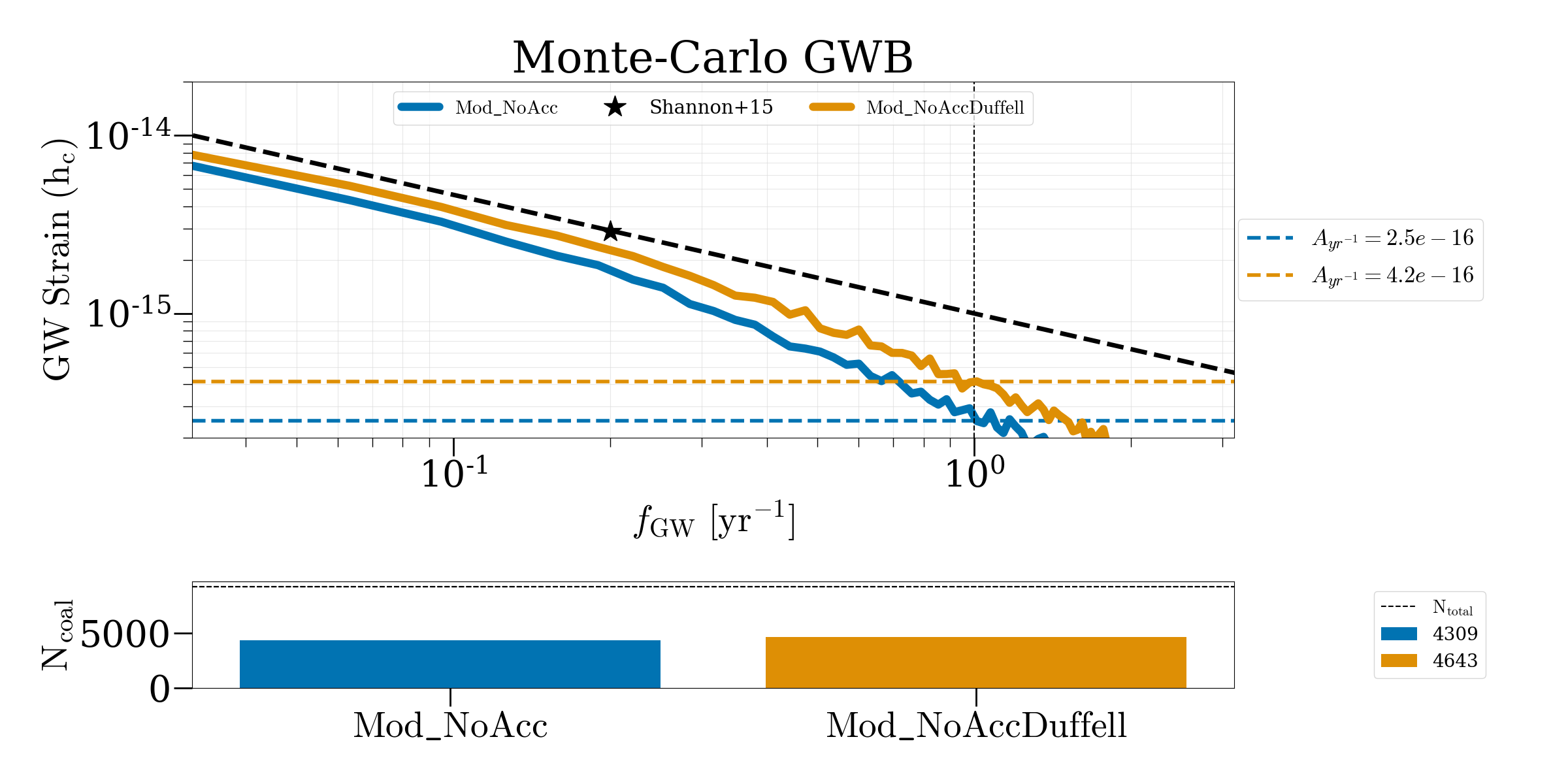}
	\caption{Top panel: MC GWB spectra comparing \texttt{Mod\_NoAcc} and \texttt{Mod\_NoAccDuffell}. The GWB amplitude in  \texttt{Mod\_NoAccDuffell} is increased by $68\%$ compared with  \texttt{Mod\_NoAcc}.\\
	Bottom panel: Number of coalescences in our MBHB sample per model. We find that out of \nsources \ systems, between  46.5\% (\texttt{Mod\_NoAcc}) and 50.0\% (\texttt{Mod\_NoAccDuffell}) of MBHBs coalesce in a Hubble time.}
	\label{fig:gwb_noacc_duffell}
\end{figure}

\label{lastpage}
\end{document}